\documentclass[osajnl,twocolumn,showpacs,superscriptaddress,10pt]{revtex4-1} 

\usepackage{graphicx}
\usepackage{dcolumn}
\usepackage{bm}
\usepackage{amsmath,amssymb}
\usepackage{comment}

\def\U#1{{%
\def\O{\mbox{O}}
\def\u{\mbox{u}}
\mathcode`\u=\mu
\mathcode`\O=\Omega
\mathrm{#1}}}

\def\ii{{\mathrm{i}}}

\def\sub#1{_{\scriptsize\mbox{#1}}}

\def\degree{\mbox{$^\circ$}}
\def\vct#1{{\mathchoice{\mbox{\boldmath$#1$}}{\mbox{\boldmath$#1$}}%
  {\mbox{\scriptsize\boldmath$#1$}}{\mbox{\scriptsize\boldmath$#1$}}}}

\begin{document}

\title{Design and analysis of frequency-independent reflectionless single-layer metafilms}

\author{Yasuhiro Tamayama}
\email{tamayama@vos.nagaokaut.ac.jp}
\affiliation{Department of Electrical, Electronics and information Engineering, Nagaoka University of
Technology, 1603-1 Kamitomioka, Nagaoka, Niigata 940-2188, Japan}
\date{\today}

 \begin{abstract}
We develop a theory for realizing frequency-independent, reflectionless, single-layer metafilms based on the Brewster effect. A metafilm designed based on the theory is numerically analyzed using a finite-difference time-domain method. The numerical analysis demonstrates that the reflectance of the metafilm vanishes independent of the frequency and that the metafilm behaves like an all-pass filter  (with finite loss). An analysis based on an electrical circuit model of the reflectionless metafilm reveals that the energy of the suppressed reflection wave is not stored in the metafilm but is radiated to the transmission direction. 
 \end{abstract}
 
\ocis{(160.3918) Metamaterials; (260.2110) Electromagnetic optics;
(160.4760) Optical properties.}

\maketitle


In recent years there has been considerable interest in controlling
electromagnetic waves using metamaterials. In particular, metafilms that
are composed of two-dimensionally arranged meta-atoms have attracted
much attention, because they can be fabricated more easily than
three-dimensional metamaterials and enable the realization of a wide
variety of elements for controlling electromagnetic waves. For example,
metafilms have been applied to polarization control~[1-5], harmonic
generation enhancement~[6-9], and slow group velocity
propagation~[10-13]. In these studies, metafilms with planar structures,
which are usually referred to as metasurfaces, were used. Although
metasurfaces are very useful for controlling electromagnetic waves,
arbitrary optical properties cannot be obtained using single-layer
metasurfaces with negligible thicknesses. This is because only electric
dipoles can be induced in the plane of metasurfaces, which sets some
restrictions on the optical properties of single-layer metasurfaces. For
example, the absorbance cannot exceed 0.5~[14,15], and the amplitude and
phase of the transmitted (reflected) wave cannot be controlled
independently~[16].

The above restrictions can be removed in multilayer metasurfaces and
metafilms with finite thicknesses. Perfect absorption~[17] and
independent control of the reflection amplitude and phase~[18]
have been realized using ground planes. In addition, control of the
transmission phase with nearly unity
amplitude~[16, 19-23]
and ground-plane-less perfect absorbers~[24, 25] have
been realized in metasurfaces (metafilms) composed of more than
two-layer structures~[16, 19, 20],
Huygens'
metasurfaces~[21-24], and
omega metafilms~[25]. Although suppression of the
transmission/reflection is essential in these examples, it is much more
difficult to suppress the reflection independent of the frequency than
the transmission. Thus far, the frequency-independent suppression of
reflection in metasurfaces (metafilms) has been realized only in
Huygens' metasurfaces~[23, 24] and omega metafilms~[25], where the amplitudes of radiated waves from the induced electric and magnetic dipoles are designed to be equal over a broad frequency range. 

In this study, we propose and numerically verify a method for suppressing reflection
independent of frequency in single-layer metafilms composed of meta-atoms with only an electric response. The frequency-independent reflectionless metafilm is designed based on the Brewster effect in metafilms~[26] and the optical properties of the designed metafilm are numerically analyzed using a finite-difference time-domain method~[27]. The reflectionless metafilm is modeled as an electrical circuit using the results of the numerical simulation, and the mechanism of the electromagnetic response of the metafilm is clarified using the electrical circuit model.


First, we briefly review the Brewster effect, the physical origin of
which is the key of the present study.
When an electromagnetic wave with certain polarization is incident on a
planar interface between two distinct media at a
particular angle, called the Brewster angle, the wave is transmitted
with zero reflection~[28]. The physical origin of the Brewster effect can be understood by considering radiations from the electric and magnetic dipoles induced in the media~[29]. Here, an electromagnetic wave is assumed to be incident from vacuum onto a purely dielectric medium for simplicity. In this case, the reflected wave consists of radiation from the electric dipoles induced in the dielectric medium. Because these electric dipoles cannot radiate electromagnetic waves in the direction of the oscillation, the reflected wave vanishes if the propagation direction of the reflected wave coincides with the oscillation direction of the electric dipoles. 

\begin{figure}[tb]
   \begin{center}
    \includegraphics[scale=0.7]{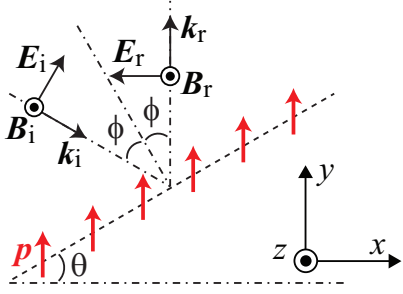}
    \caption{Geometry of coordinate system considered in the theory. 
    The oscillation direction of the induced electric
    dipoles $\vct{p}$ in the metafilm is represented by red bold arrows and
    the surface of the metafilm is shown by the dashed line. The
    subscripts i and r denote the incident and reflected waves,
    respectively.}
    \label{fig:theory}
   \end{center}
\end{figure}

We develop a method for suppressing the reflection from single-layer
metafilms independent of the incident frequency based on the Brewster
effect. Let us assume that the oscillation direction of the electric
dipoles in a metafilm is in the $y$ direction and that meta-atoms are
arranged at angle $\theta$ with respect to the $x$ axis, as shown in
Fig.\,\ref{fig:theory}. (The system is assumed to be periodic in the $z$
direction.) The incident and reflection angles are denoted by $\phi$. A
geometric calculation shows that the reflected wave propagates along the
$y$ direction when $\theta = \phi$ is satisfied. Thus, the reflected
wave vanishes if the arrangement of the meta-atoms and the propagation
direction of the incident wave are chosen such that $\theta = \phi$ is
satisfied. The reflection can be suppressed as long as the electric
dipoles oscillate along the $y$ direction; therefore, the suppression of the reflection can be achieved independent of the incident frequency.


\begin{figure}[tb]
 \begin{center}
  \includegraphics[scale=0.8]{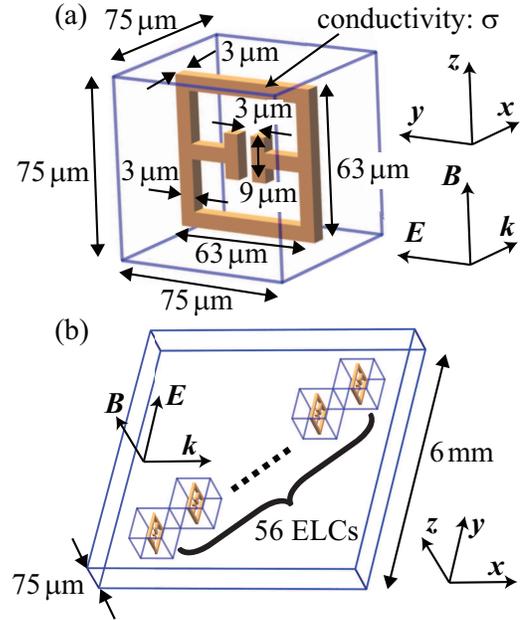}
  \caption{Geometrical parameters of (a) unit cell of the metafilm and
  (b) the simulation system. }
  \label{fig:sim_sys}
 \end{center}
\end{figure}

We design a frequency-independent, reflectionless, single-layer metafilm
based on the above theory. In this study, an electric-field-coupled
inductive-capacitive resonator (ELC)~[30] shown in
Fig.\,\ref{fig:sim_sys}(a) is used as the meta-atom. (Other
structures with a purely dielectric response can also be used as the
meta-atom.) This structure is
one of the commonly used meta-atoms and exhibits a purely dielectric
response for electromagnetic waves that propagate along the $x$ direction.
When a $y$-polarized electromagnetic wave is incident on the ELC, an electric dipole is induced in the $y$ direction. The frequency dependence of the ELC response is Lorentz-type with a resonance frequency that is determined by the inductance of the ring and the capacitance of the central gap. The metafilm is constructed by periodically arranging ELCs in the $z$ direction and in the direction of $\theta = \phi = \pi / 4$, for which the interaction between the incident wave and the metafilm becomes strongest because the propagation direction of the incident wave coincides with the direction in which the radiation from each electric dipole is largest. Such a structure can be fabricated using microelectromechanical systems-based technologies~[31]. 

We numerically analyzed the transmission and reflection properties of the
designed metafilm using a finite-difference time-domain method. Figure
\ref{fig:sim_sys}(b) shows the simulation system. The dimension of the
simulation space was $6\,\U{mm} \times 6\,\U{mm} \times
75\,\mu\U{m}$. Periodic boundary conditions were applied to the $z$
direction and perfectly matched layer boundary conditions were applied to
the $x$ and $y$ directions. The unit cell of the metafilm, shown in
Fig.\,\ref{fig:sim_sys}(a), was periodically arranged in the direction of
$\theta = \pi /4$, and the number of the arranged ELCs was 56. A
$y$-polarized Gaussian beam with a focal spot width of $1.2\,\U{mm}$ and
a Rayleigh range of $3.8\,\U{mm}$ was incident from the $-x$ direction
onto the metafilm at the incident angle of $\phi = \pi /4$. The
transmittance (reflectance) of the metafilm is calculated as the ratio
of the amplitude of the transmitted (reflected) wave to the amplitude of
the incident wave in the far-field region. 

\begin{figure*}[tb]
 \begin{center}
  \includegraphics[scale=0.7]{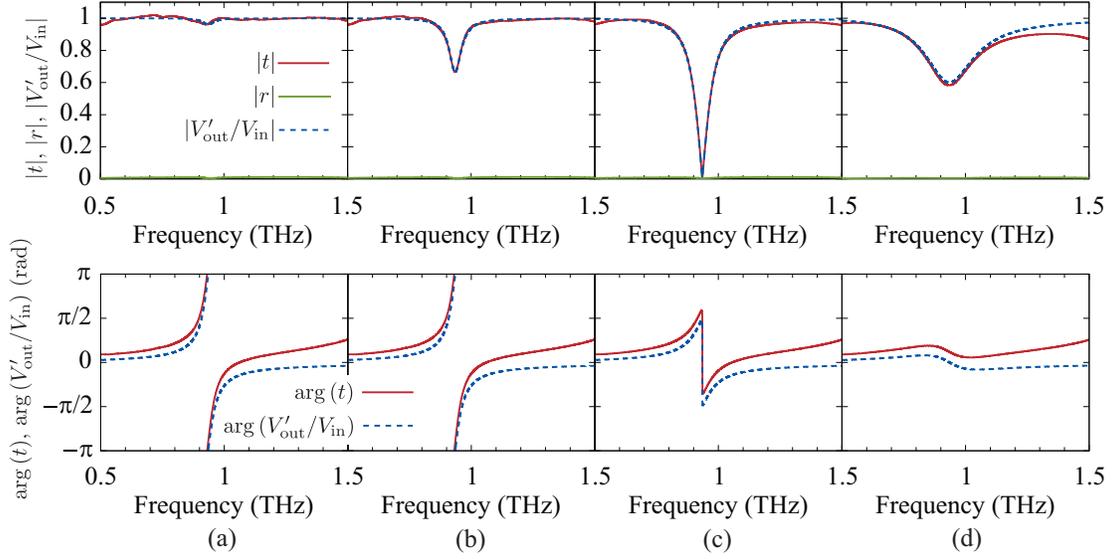}
  \caption{(Solid curves) Frequency dependences of (upper row) the amplitude transmittance
  $|t|$ and amplitude reflectance $|r|$, and (lower
  row) the transmission phase $\arg{(t)}$ of the metafilm when the conductivity $\sigma$
  of the ELC is
  (a) $1.0\times 10^7 \,\U{S/m}$, (b) $1.0\times 10^6\,\U{S/m}$,
  (c) $2.0\times 10^5 \,\U{S/m}$, and (d) $5.0\times 10^4\,\U{S/m}$.
  (Dashed curves) Frequency dependences of (upper row) the absolute value
  and (lower row) the argument of $V\sub{out}^{\prime} / V\sub{in}$ in the
  electrical circuit model when (a) $R\sub{nr} / R\sub{r} = 0.02$, (b)
  $0.2$, (c) $1.001$, and (d) $4$.
  The values of $R\sub{nr} / R\sub{r}$ in these four conditions
  are determined by assuming that
  $R\sub{nr}$ for $\sigma = 2.0\times 10^5 \,\U{S/m}$ is equal to
  $R\sub{r}$. However,  
  $R\sub{nr}$ for $\sigma = 2.0\times 10^5 \,\U{S/m}$
  is not exactly equal to $R\sub{r}$; thus, the value of
  $R\sub{nr} / R\sub{r}$ in condition (c) is slightly shifted from $1$
  so that $V\sub{out}^{\prime} / V\sub{in}$ well agrees with $t$.
  The other electrical circuit parameters used in this calculation are
  $L=\beta \times 1.702 \times 10^{-13} \,\U{H}$, $C=(1/\beta) \times
  1.702 \times 10^{-13} \,\U{F}$, and $R\sub{r}=\beta \times
  (1/17)\,\Omega$, where $\beta$ ($>0$) is a dimensionless constant and does not
  affect $V\sub{out}^{\prime}$.}
  \label{fig:spectra}
 \end{center}
\end{figure*}

Figure \ref{fig:spectra} shows the transmission and reflection spectra
of the metafilm when the conductivity $\sigma$ of the ELC is equal to
$1.0\times 10^7 \,\U{S/m}$, $1.0\times 10^6\,\U{S/m}$, $2.0\times 10^5
\,\U{S/m}$, and $5.0\times 10^4\,\U{S/m}$. To allow for a detailed
discussion of the electromagnetic response of the metafilm in the
following paragraphs, the optical properties are calculated for several
values of $\sigma$. The reflectance vanishes independent of the incident
frequency in every case. Therefore, the theory for realizing
frequency-independent reflectionless single-layer metafilms is
confirmed, and hereafter, we focus on the transmission properties of the
reflectionless metafilms. The amplitude transmittance is almost unity
independent of the frequency for $\sigma = 1.0\times 10^7
\,\U{S/m}$. This observation can be understood from energy
conservation. The absorption in the metafilm is negligible because the
ELC is made of an almost perfect electric conductor. Both the reflection
and absorption vanish in the metafilm, resulting in perfect
transmission. The transmittance at the resonance frequency,
$0.935\,\U{THz}$, decreases as the conductivity decreases, and it
vanishes for $\sigma = 2.0\times 10^5\,\U{S/m}$, which implies that
perfect absorption takes place. The transmittance increases with further
decreasing the conductivity. The transmission phase monotonically
increases with the incident frequency and ranges from $-\pi$ to $\pi$
for $\sigma > 2 \times 10^5\,\U{S/m}$. On the other hand, for $\sigma
< 2 \times 10^5\,\U{S/m}$, the transmission phase decreases with
increasing the incident frequency at around the resonance frequency and
takes on a narrower range of values compared with the case of large conductivity. Although the reflectionless metafilm is composed of a commonly used Lorentz-type meta-atom, the dependence of the metafilm transmittance on the frequency and the loss in meta-atoms is quite different from that of usual Lorentz-type metamaterials.


\begin{figure}[tb]
 \begin{center}
  \includegraphics[scale=1.2]{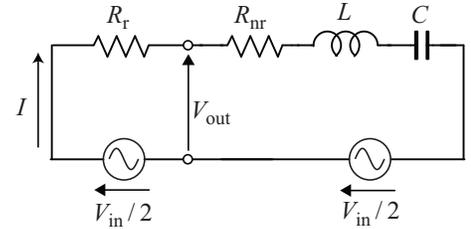}
  \caption{Electrical circuit model of the frequency-independent,
   reflectionless metafilms. $V\sub{out} = V^{\prime}\sub{out}/2$. }
  \label{fig:all-pass}
 \end{center}
\end{figure}

Next, we create and analyze an electrical circuit model of the
frequency-independent reflectionless metafilm to understand its
electromagnetic response mechanism. Since the metafilm behaves like an
all-pass filter when the ELC is made of a perfect electric conductor, we
consider the electrical circuit shown in Fig.\,\ref{fig:all-pass}, where
$L$, $C$, $R\sub{r}$, and $R\sub{nr}$ represent the inductance of the
ring, capacitance at the central gap, radiation loss, and non-radiation
loss of the ELC, respectively. The voltage source $V\sub{in}$
corresponds to the incident wave. Applying Kirchhoff's voltage law to
the electrical circuit yields $V\sub{out}=-\{ [R\sub{r}-R\sub{nr}+\ii
\omega L + (\ii \omega C)^{-1}]/[R\sub{r}+R\sub{nr}-\ii \omega L - (\ii
\omega C)^{-1}] \} (V\sub{in}/2)$. In the case of $R\sub{nr}=0$,
$|V\sub{out}/V\sub{in}|=1/2$ is satisfied independent of the source
frequency and $\arg(V\sub{out}/V\sub{in})$ ranges from $-\pi$ to
$\pi$. Assuming that $V\sub{out}^{\prime} = 2V\sub{out}$ is the output
and $V\sub{in}$ is the input of the electrical circuit, the electrical
circuit can be regarded as an all-pass filter with unity gain. Figure
\ref{fig:spectra} shows the frequency dependence of
$V\sub{out}^{\prime}/V\sub{in}$ for four different values of
$R\sub{nr}/R\sub{r}$, which correspond to the four conditions in the
numerical simulation. The frequency dependences of the absolute value and
argument of $V\sub{out}^{\prime} / V\sub{in}$ agree well with the
transmission properties of the reflectionless metafilm. Therefore, the
electrical circuit shown in Fig.\,\ref{fig:all-pass} can safely be
regarded as a suitable electrical circuit model of the reflectionless metafilm. It is important to notice that $V\sub{out}^{\prime} = V\sub{in} - 2R\sub{r}I$ is satisfied in this electrical circuit. The mechanism of the electromagnetic response of the metafilm can be understood based on this equation.

To understand the meaning of the above equation, we consider the
electromagnetic response of the metafilm with $\theta = 90\degree$
(metasurface) for normal incidence. In this case, the amplitudes of the
radiated waves from the metasurface propagating in the transmission and
reflection directions are the same. Although the configuration of the
electrical circuit model of the metasurface is identical to that of the
electrical circuit shown in Fig.\,\ref{fig:all-pass}, the transmittance
does not correspond to $V\sub{out}^{\prime} / V\sub{in}$ in this
case. The amplitude of the transmitted wave corresponds to
$V\sub{in}-R\sub{r}I$. This is because the transmitted wave is composed
of a superposition of the incident wave and the radiation from the
metasurface, which is proportional to the current $I$ and the radiation
loss $R\sub{r}$ in the meta-atom, and the transmittance of the
Lorentz-type metasurface at the resonance frequency vanishes in the case
of  $R\sub{nr}=0$~[4, 32]. Therefore, the amplitude of the radiated wave from the metasurface propagating in the transmission direction corresponds to $R\sub{r}I$.

Now we discuss the meaning of the equation, $V\sub{out}^{\prime} =
V\sub{in} - 2R\sub{r}I$. Based on the above analysis and equation, the
amplitude of the radiated wave from the reflectionless metafilm
propagating in the transmission direction is found to be twice as large
as that of the radiated wave from the metasurface propagating in the
transmission direction. This implies that the energy of the suppressed
reflection wave is not stored in the reflectionless 
metafilm but is radiated to the transmission direction.

For further understanding the physical mechanism of the electromagnetic
response, we consider the transmittance of the metafilm at the resonance
frequency for two special cases, $R\sub{nr}=0$ and $R\sub{nr}=R\sub{r}$,
using the result of the above discussion. When $R\sub{nr}=0$, the metafilm radiates
an electromagnetic wave with two times larger amplitude than and
opposite phase to the incident wave. This implies that the absolute
value and argument of the transmittance are unity and $\pi$,
respectively. When $R\sub{nr}=R\sub{r}$, the current $I$ that flows in the
meta-atom is equal to half of that in the case of $R\sub{nr}=0$. Thus,
the metafilm radiates an electromagnetic wave with the same amplitude as
and opposite phase to the incident wave, and the transmittance
vanishes. Here, the radiated power from the metafilm is equal to the
absorbed power in the metafilm because $R\sub{nr}=R\sub{r}$
is satisfied. (This condition is known as critical coupling~[33].)
Therefore, perfect absorption occurs. 

Note that the macroscopic optical property of the reflectionless
metafilm is similar to that of a Huygens' metasurface with spectrally
overlapping electric and magnetic resonances~[34]. Although only
electric dipoles are induced in the reflectionless metafilm, the
suppression of the reflection causes the equivalent effect for the
radiation from magnetic dipoles, which has the same amplitude as that
from the induced electric dipoles.


In conclusion, we developed a theory for realizing frequency-independent, reflectionless, single-layer metafilms and analyzed the optical properties of one such metafilm, which was designed based on the theory. The frequency-independent suppression of the reflection is realized by arranging the meta-atoms so that the oscillation direction of the electric dipole induced in the meta-atom coincides with the propagation direction of the reflected wave. The designed metafilm was numerically analyzed using a finite-difference time-domain method. The reflectance of the metafilm was confirmed to vanish independent of the incident frequency and of the loss in the meta-atom. An electrical circuit model of the reflectionless metafilm was created and the mechanism of the electromagnetic response of the metafilm was analyzed based on the model. The analysis revealed that the energy of the suppressed reflection wave is not stored in the metafilm but is radiated to the transmission direction.

The frequency-independent reflectionless metafilm behaves as an all-pass
filter when the non-radiative loss $R\sub{nr}$ is much smaller than the
radiation loss $R\sub{r}$. All-pass filters are important elements for
applications such as signal processing~[35, 36]. When
$R\sub{nr}=R\sub{r}$ is satisfied, perfect absorption can be realized
using the single-layer metafilm developed here. The suppression of the
reflection in our method is realized as long as the induced electric
dipole oscillates along the $y$ direction. Thus, transmission-type spatial phase modulators and multiband perfect absorbers may be realized using our method. The concept of suppressing reflection by considering the radiation pattern of meta-atoms and the propagation direction of the reflected wave may yield various useful devices for controlling electromagnetic waves.

\mbox{}


\noindent
{\normalsize{\bf REFERENCES}}
\begin{enumerate}
\setlength{\parskip}{0cm}
\setlength{\itemsep}{0cm} 
\renewcommand{\labelenumi}{\arabic{enumi}. }
{\small 

\item
J.~Hao, Y.~Yuan, L.~Ran, T.~Jiang, J.~A. Kong, C.~T. Chan, and L.~Zhou, Phys.
  Rev. Lett. \textbf{99}, 063908 (2007).

\item
N.~Yu, F.~Aieta, P.~Genevet, M.~A. Kats, Z.~Gaburro, and F.~Capasso, Nano Lett.
  \textbf{12}, 6328 (2012).

\item
N.~K. Grady, J.~E. Heyes, D.~R. Chowdhury, Y.~Zeng, M.~T. Reiten, A.~K. Azad,
  A.~J. Taylor, D.~A.~R. Dalvit, and H.-T. Chen, Science \textbf{340}, 1304
  (2013).

\item
S.-C. Jiang, X.~Xiong, P.~Sarriugarte, S.-W. Jiang, X.-B. Yin, Y.~Wang, R.-W.
  Peng, D.~Wu, R.~Hillenbrand, X.~Zhang, and M.~Wang, Phys. Rev. B \textbf{88},
  161104 (2013).

\item
Y.~Ke, Y.~Liu, Y.~He, J.~Zhou, H.~Luo, and S.~Wen, Appl. Phys. Lett. \textbf{107},
  041107 (2015).

\item
M.~W. Klein, C.~Enkrich, M.~Wegener, and S.~Linden, Science \textbf{313}, 502
  (2006).

\item
E.~Kim, F.~Wang, W.~Wu, Z.~Yu, and Y.~R. Shen, Phys. Rev. B \textbf{78}, 113102
  (2008).

\item
R.~Czaplicki, H.~Husu, R.~Siikanen, J.~M\"{a}kitalo, M.~Kauranen, J.~Laukkanen,
  J.~Lehtolahti, and M.~Kuittinen, Phys. Rev. Lett. \textbf{110}, 093902 (2013).

\item
K.~O'Brien, H.~Suchowski, J.~Rho, A.~Salandrino, B.~Kante, X.~Yin, and
  X.~Zhang, Nature Mater. \textbf{14}, 379 (2015).

\item
P.~Tassin, L.~Zhang, T.~Koschny, E.~N. Economou, and C.~M. Soukoulis, Phys. Rev.
  Lett. \textbf{102}, 053901 (2009).

\item
C.~Kurter, P.~Tassin, L.~Zhang, T.~Koschny, A.~P. Zhuravel, A.~V. Ustinov,
  S.~M. Anlage, and C.~M. Soukoulis, Phys. Rev. Lett. \textbf{107}, 043901
  (2011).

\item
Y.~Tamayama, T.~Nakanishi, and M.~Kitano, Phys. Rev. B \textbf{85}, 073102
  (2012).

\item
Y.~Tamayama, K.~Hamada, and K.~Yasui, Phys. Rev. B \textbf{92}, 125124 (2015).

\item
D.~Pozar, IEEE Antennas Propag. Mag. \textbf{46}, 144 (2004).

\item
S.~Thongrattanasiri, F.~H.~L. Koppens, and F.~J.
  \mbox{Garc\'{\i}a~de~Abajo}, Phys. Rev. Lett. \textbf{108}, 047401 (2012).

\item
F.~Monticone, N.~M. Estakhri, and A.~Al\`{u}, Phys. Rev. Lett. \textbf{110},
  203903 (2013).

\item
C.~M. Watts, X.~Liu, and W.~J. Padilla, Adv. Opt. Mater. \textbf{24}, OP98
  (2012).

\item
M.~Kim, A.~M.~H. Wong, and G.~V. Eleftheriades, Phys. Rev. X \textbf{4}, 041042
  (2014).

\item
C.~Pfeiffer and A.~Grbic, Phys. Rev. Appl. \textbf{2}, 044012 (2014).

\item
J.~Luo, H.~Yu, M.~Song, and Z.~Zhang, Opt. Lett. \textbf{39}, 2229 (2014).

\item
C.~Pfeiffer and A.~Grbic, Phys. Rev. Lett. \textbf{110}, 197401 (2013).

\item
M.~Selvanayagam and G.~V. Eleftheriades, Opt. Express \textbf{21}, 14409 (2013).

\item
K.~E. Chong, I.~Staude, A.~James, J.~Dominguez, S.~Liu, S.~Campione, G.~S.
  Subramania, T.~S. Luk, M.~Decker, D.~N. Neshev, I.~Brener, and Y.~S.
  Kivshar, Nano Lett. \textbf{15}, 5369 (2015).

\item
V.~S. Asadchy, I.~A. Faniayeu, Y.~Ra'di, S.~A. Khakhomov, I.~V. Semchenko, and
  S.~A. Tretyakov, Phys. Rev. X \textbf{5}, 031005 (2015).

\item
A.~Balmakou, M.~Podalov, S.~Khakhomov, D.~Stavenga, and I.~Semchenko, Opt. Lett.
  \textbf{40}, 2084 (2015).

\item
Y.~Tamayama, Opt. Lett. \textbf{40}, 1382 (2015).

\item
A.~Taflove and S.~C. Hagness, \emph{Computational Electrodynamics: The
  Finite-Difference Time-Domain Method} (Artech House, Norwood, MA, 2005), 3rd
  ed.

\item
B.~E.~A. Saleh and M.~C. Teich, \emph{Fundamentals of Photonics}
  (Wiley-Interscience, Hoboken, NJ, 2007), 2nd ed.

\item
G.~P. Sastry and S.~Chakrabarty, Eur. J. Phys. \textbf{8}, 125 (1987).

\item
D.~Schurig, J.~J. Mock, and D.~R. Smith, Appl. Phys. Lett. \textbf{88}, 041109
  (2006).

\item
H.~Tao, A.~C. Strikwerda, K.~Fan, W.~J. Padilla, X.~Zhang, and R.~D.
  Averitt, Phys. Rev. Lett. \textbf{103}, 147401 (2009).

\item
Y.~Tamayama, K.~Yasui, T.~Nakanishi, and M.~Kitano, Appl. Phys. Lett.
  \textbf{105}, 021110 (2014).

\item
Z.~Ruan and S.~Fan, Phys. Rev. Lett. \textbf{105}, 013901 (2010).
     
\item
M.~Decker, I.~Staude, M.~Falkner, J.~Dominguez, D.~N. Neshev, I.~Brener,
  T.~Pertsch, and Y.~S. Kivshar, Adv. Opt. Mater. \textbf{3}, 813 (2015).

\item
Y.~Horii, S.~Gupta, B.~Nikfal, and C.~Caloz, IEEE Microwave Wireless Compon.
  Lett., \textbf{22}, 1 (2012).

\item
K.~Goda and B.~Jalali, Nature Photon. \textbf{7}, 102 (2013).
	
}

\end{enumerate}


\end{document}